\documentclass[aps,showpacs,floatfix,twocolumn]{revtex4}
\usepackage{graphicx}
\makeatletter
\newbox\slashbox \setbox\slashbox=\hbox{\large$/$}
\def\pslash#1{\setbox\@tempboxa=\hbox{$#1$}
  \@tempdima=0.5\wd\slashbox \advance\@tempdima 0.5\wd\@tempboxa
  \copy\slashbox \kern-\@tempdima \box\@tempboxa}

\makeatother
\newcommand{\mat}{\left ( \begin{array}{cc}}
\newcommand{\emat}{\end{array} \right )}
\newcommand{\be}{\begin{eqnarray}}
\newcommand{\ee}{\end{eqnarray}}
\newcommand{\ba}{\begin{array}}
\newcommand{\ea}{\end{array}}

\def\beq{\begin{equation}}
\def\eeq{\end{equation}}

\begin{document}
\title{Classical singularities and Semi-Poisson statistics\\
in quantum chaos and disordered systems}
\author{Antonio M. Garc\'{\i}a-Garc\'{\i}a}
\affiliation{Physics Department, Princeton University, Princeton, New Jersey 08544, USA}
\begin{abstract}
We investigate a 1D disordered Hamiltonian with a non analytical step-like dispersion 
 relation whose 
level statistics is exactly described by Semi-Poisson statistics(SP). It is shown that 
this result is robust, namely, does not depend neither on the microscopic details of the potential nor 
 on a magnetic flux but only on the type of non-analyticity. We also argue 
 that a deterministic kicked rotator with a non-analytical step-like potential has the same 
 spectral properties. 
Semi-Poisson statistics (SP), typical of pseudo-integrable billiards, has been frequently
claimed to describe critical statistics, namely, the level statistics of a disordered system 
at the Anderson transition 
(AT). However we provide convincing evidence they are indeed different: each of them has its origin
 in a different type of classical singularities.

\end{abstract}
\maketitle
The properties of a quantum particle in a random potential 
  is one of the  
 most intensively studied problems in condensed matter physics since the landmark paper 
by Anderson \cite{anderson}. 
According to the one parameter scaling theory, in more 
 than two dimensions, there exists 
a metal insulator transition for   
 a critical amount of disorder. Unfortunately the AT
 in three and higher dimensions takes place in a region of strong
 disorder not directly accessible to current analytical techniques.
Despite the lack of rigorous analytical results,
it is by now well established, mainly through numerical simulations, 
that the AT is fully characterized 
 by the level statistics and the anomalous scaling of the eigenfunction moments,
 ${\cal P}_q=\int d^dr |\psi({\bf r})|^{2q} \propto L^{-D_q(q-1)}$ 
 with respect to the sample size $L$, where $D_q$ is a set of exponents describing the AT.  
Eigenfunctions with such a nontrivial (multi) scaling are usually dubbed multifractals 
\cite{aoki}. 
Critical statistics \cite{sko,kravtsov97}, the level statistics at the AT, 
is intermediate between 
 Wigner-Dyson (WD) and Poisson statistics.
Typical features include: scale invariant spectrum \cite{sko},
 level repulsion and asymptotically linear number 
 variance \cite{chi}. 

Both level statistics and multifractal properties are universal 
 in the sense that parameters such as the slope of the number variance or the set 
of multifractals exponents $D_q$ depend only on the dimension 
 of the system and not on boundary conditions,
 shape of the system, or the microscopic details of the disordered potential. 
However the functional form of level correlators as the level spacing distribution
or the number variance may be affected by such variables. 

A natural question to ask 
 is whether critical statistics can be reproduced by generalized random matrix models (RMM). 
The answer to this question is positive: Critical statistics has been found 
 in RMM
 based on soft confining potentials 
 \cite{log}, effective eigenvalue distributions \cite{Moshe,ant4} related to the Calogero-Sutherland model at finite temperature
 and random banded matrices with power-law decay \cite{ever}. 
 The latter is specially interesting since an AT has been analytically established
by mapping the problem onto a non linear $\sigma$ model.
In a certain region of parameters, all of above RMM share the same spectral kernel,
$K(s)=T\frac{\sin(\pi s)}{\sinh(\pi sT)}$ ($T$ is a free parameter which enters in the definition of the above models). 

In the context of quantum chaos similarities with an AT have also been
found in a variety of systems:  
Coulomb billiard\cite{altshu}, Anisotropic Kepler problem \cite{wintgen} and 
generalized Kicked rotors \cite{bao}. In all of them the classical potential has 
a singularity and the classical dynamics is intermediate between integrable and chaotic. 
In a recent letter \cite{ant9} we have put forward a new universality class in quantum chaos based 
 precisely on the relation between the 
 type of singularity of the classical potential and the properties of the 
 quantum eigenstates. Specifically, for a certain kind of non-analyticity, it was found that
 the level statistics is described by critical statistics and the eigenstates are multifractal as at the AT. 

Similar properties have also been found in  pseudointegrable billiards \cite{bogo3,bogo04}. 
The level statistics of these models is very well described by 
 a phenomenological short range plasma model \cite{bogo3} 
  whose joint distribution of eigenvalues is given by 
the classical Dyson gas with the logarithmic 
pairwise interaction restricted to a finite number $k$ of nearest neighbors. 
 Explicit expressions for the level statistics, usually referred to as SP statistics, are available for general $k$. For $k=2$,
 $R_2(s)=1-e^{-4s}$, $P(s)=4s e^{-2s}$ and $\Sigma^2(L) = L/2 +(1-e^{-4L})/8$ where $R_2(s)=- K(s)^2$ is the two 
level correlation function, $P(s)$ is the level spacing distribution and $\Sigma^2(L)$ is the number variance.
Thus SP statistics
reproduces typical characteristics of critical statistics 
as level repulsion, linear number variance, with a slope depending on $k$.

The aim of this paper is to incorporate the systems described by 
SP into the classification introduced in \cite{ant9} which relate quantum properties 
with singularities of the classical potential. We shall see that differences between  SP  and
 critical statistical stem from the fact that both come from different classical singularities.
 In order to proceed we propose
a non-analytic Hamiltonian whose level statistics is exactly described 
by semi-Poisson statistics.
It is then shown that differences between critical and semi-Poisson statistics 
 are due to the fact that both come from different type of singularity in the classical potential.         
  \begin{figure}[ht]
\includegraphics[width=0.9\columnwidth,clip,angle=0]{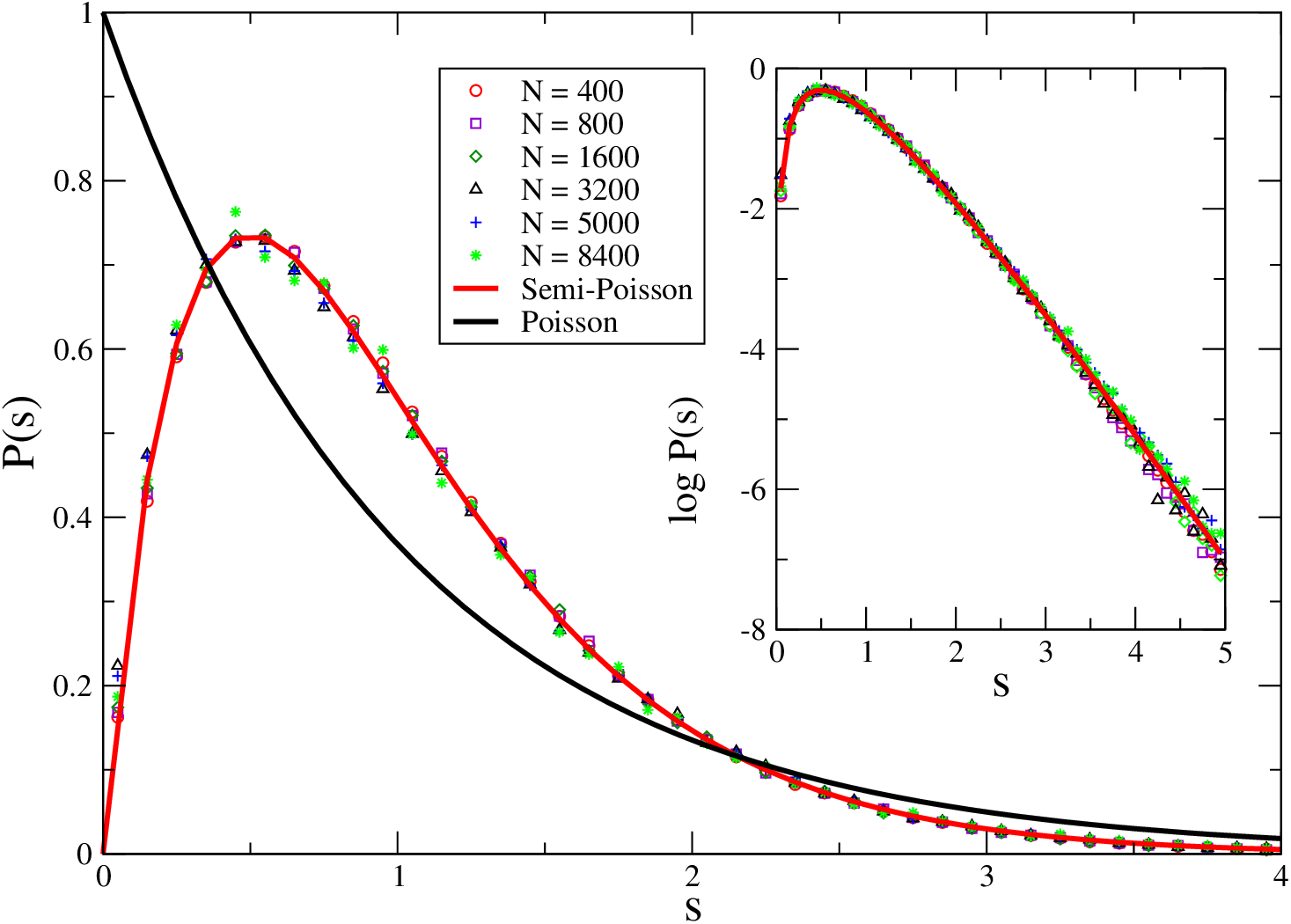}
\vspace{3mm}
\caption{Level spacing distribution for $\gamma = \pi/2$ and $A = 10$ and 
different system sizes. The agreement with SP is impressive even in the tail of $P(s)$
(see Inset).}  \label{fig1} \end{figure}

{\it The models--}We start our investigation with a generalized kicked rotor with a a step-like singularity,
\be
\label{ourmodel}
{\cal H}= \frac{p^2}2 +V(q)\sum_n\delta(t -nT)
\ee
with $V(q)$ a step function, $V(q)= \alpha$; $q\in [-\beta,\beta)$ and otherwise $V(q) = 0$ ; 
$\beta \in [-\pi,\pi[$ and $\alpha$ is a real number.
By using the method introduced in \cite{fishman} one can map
the evolution matrix associated to Eq. (\ref{ourmodel}) onto 
 the following 1D Anderson model with long-range hopping. 
\be
\label{ourmodel1}
{\cal H} \psi_i= \epsilon_i \psi_i + \sum_j F(i-j)\psi_j
\ee
where $\epsilon_i \sim \tan(\alpha i^2)$ are pseudo random numbers (for $\alpha > 1$) and \cite{fishman} $F(m-n) = 
\int_{-\pi}^{-\pi}d\theta \tan(V(\theta))e^{-i\theta(m-n)} =
A\frac{\sin \gamma (m-n)}{m-n}$ with $A,W$  
real positive constants related to $\beta, \alpha$. The case $\gamma = \pi/2$ describes the high energy limit of the interaction
 of a charged particle with quenched Coulomb scatterers of alternate sign. 
 
In this paper we shall mainly investigate the Hamiltonian Eq. (\ref{ourmodel1}) assuming that 
 $\epsilon_i$ is a random
 number extracted from a box  distribution $[-W/2,W/2]$. 
 We proceed so in order to make an accurate analysis of the level statistics necessary for
 a precise comparison with SP statistics. 
 However a detailed study of the Hamiltonian  Eq. \ref{ourmodel},  
 including the details of the multifractal spectrum and the quantum wave-packet evolution, will be published elsewhere \cite{ant10}.  

{\it Results--}

We  first state our main 
results:\\
1. For $A \gg W$ and any $\gamma$ in Eq. \ref{ourmodel1}, the spectrum 
  is scale invariant and the level statistics is exactly described by 
  semi-Poisson statistics (these findings are in agreement with a
unpublished calculations of Charles Schmit) .
2. There is a transition from SP to Poisson 
   statistics as the strength of disorder is increased. 
3. In the region $A \gg W$ the 
  eigenfunctions are multifractal but with a multifractal spectrum clearly different 
  from the one observed at an AT.\\

We start by providing analytical evidence that the level statistics of Eq. (\ref{ourmodel1})  is  
 described by SP statistics.
 We express the  Hamiltonian (\ref{ourmodel}) in Fourier space as,
\be
{\cal H} =  E_k\psi_k + \sum_{k \neq k'}{\hat A}(k,k')\psi_k'
\ee
where $E_k = \sum_r \frac{\sin \gamma r}{r} e^{i k r}$ and ${\hat A}(k,k') =\frac{1}{N}
\sum_n \epsilon_n e^{-in(k-k')}$. We fix $\gamma = \pi/2$ (our findings do not depend on 
 $\gamma$), after a simple calculation we found that $E_k$ is not a smooth function (this step-like singularity
 is indeed the seed for the appearance of SP), 
$E_k = A\pi/2$ for $k < \pi$ and $E_k = -A\pi/2$ for $ k > \pi$. 
 There are thus only two possible values of the  energy separated by a gap $\delta = A\pi$. 
  Upon adding a weak ($A \gg W$) disordered potential this degeneracy is lifted 
 and the spectrum is composed  of two separate bands
 of size $\sim W$ around each of the bare points $-A\pi/2,A\pi/2$.  Since the Hamiltonian 
 is invariant under the transformation $A \rightarrow -A$, the spectrum 
must also posses that symmetry. 
 That means that, to leading order in $A$ (neglecting $1/A$ corrections), 
the number of independent eigenvalues of Eq. \ref{ourmodel1} is $n/2$
 instead of $n$.

We now show how this degeneracy affects the roots (eigenvalues) of the characteristic polynomial $P(t)= \det (H-tI)$. 
Let $P_{dis}(t)=a_0+a_1t+\ldots a_n t^n$ be the characteristic polynomial associated 
with the disordered part of the Hamiltonian. We remark that despite of its complicated, its roots, by definition, are random numbers with a box distribution $[-W,W]$. On the other hand, in the clean case $P_{clean}(t)=(t-A)^{n/2}(t+A)^{n/2}$ ($\pi$ factors are not considered). 
Due to the $A \rightarrow -A$ symmetry the 
full (Eq. \ref{ourmodel1}) case $P_{full}$ corresponds with $P_{dis}$ but replacing $t^k$ factors by a 
combination  $(t-A)^{k_1}(t+A)^{k_2}$ with $k_1 + k_2 = k$. The roots of $P_{full}$ will be 
in general complicated functions of $A$. However in the limit  of interest, $A \gg W \rightarrow \infty$, an analytical evaluation is possible. 
By setting $t = t_1 -A$ we look for roots $t_1$ of order the unity in the $A$ band.
We next perform an expansion of the characteristics polynomial $P_{full}$ to leading order in $A$.
 Thus we keep terms $A^{n/2}$ and neglect lower powers in $A$. The resulting $P_{full}$ is given by,
 $P_{full}= t_{1}^{n/2} +a_{n-2}{t_1}^{n/2-2}/3+a_{n-3}{t_1}^{n/2-3}/4+ \ldots 2 a_{n/2+1}{t_1}/n+ 2 a_{n/2}/(n+2)$ where the coefficients $a_n$  are the {\it same} than those of $P_{dia}$ above but only $n/2$ of them appear in the full case. 
The eigenvalues $\epsilon'_{i}$ of Eq.({\ref{ourmodel1}}) around the $A$ band are $\epsilon'_i = A + \beta_i$ with $\beta_i$ a root of $P_{full}$. The effect of the long range interaction is just to 
 remove all the terms with coefficients $a_0$ to $a_{n/2}$ from the characteristic polynomial of the diagonal disordered case. The spectrum is thus that of a pure diagonal disorder 
where half of the eigenvalues have been removed. The remaining eigenvalues are 
 still symmetrically distributed (the 
  ones with largest modulus are well approximated by $t_{max}= \pm \sqrt{\frac{a_{n-2}}{3}}$) around $A$. 
That means, by symmetry considerations, that the removed ones must be  either the odd or the even ones. 
This is precisely the definition of semi-Poisson statistics.
In conclusion, the power-law random banded reproduces exactly the mechanism which is utilized in the very definition
 of semi-Poisson statistics. We finally mention that the only effect of the coefficients $3,4\ldots, n+2/2$ is to renormalize the effective size of the spectrum: $\sim 2W$ for diagonal disorder and $\sim 2W/\sqrt{3}$
 for the Eq. (\ref{ourmodel1}). 
\begin{figure}[ht]

\includegraphics[width=0.7\columnwidth,clip,angle=0]{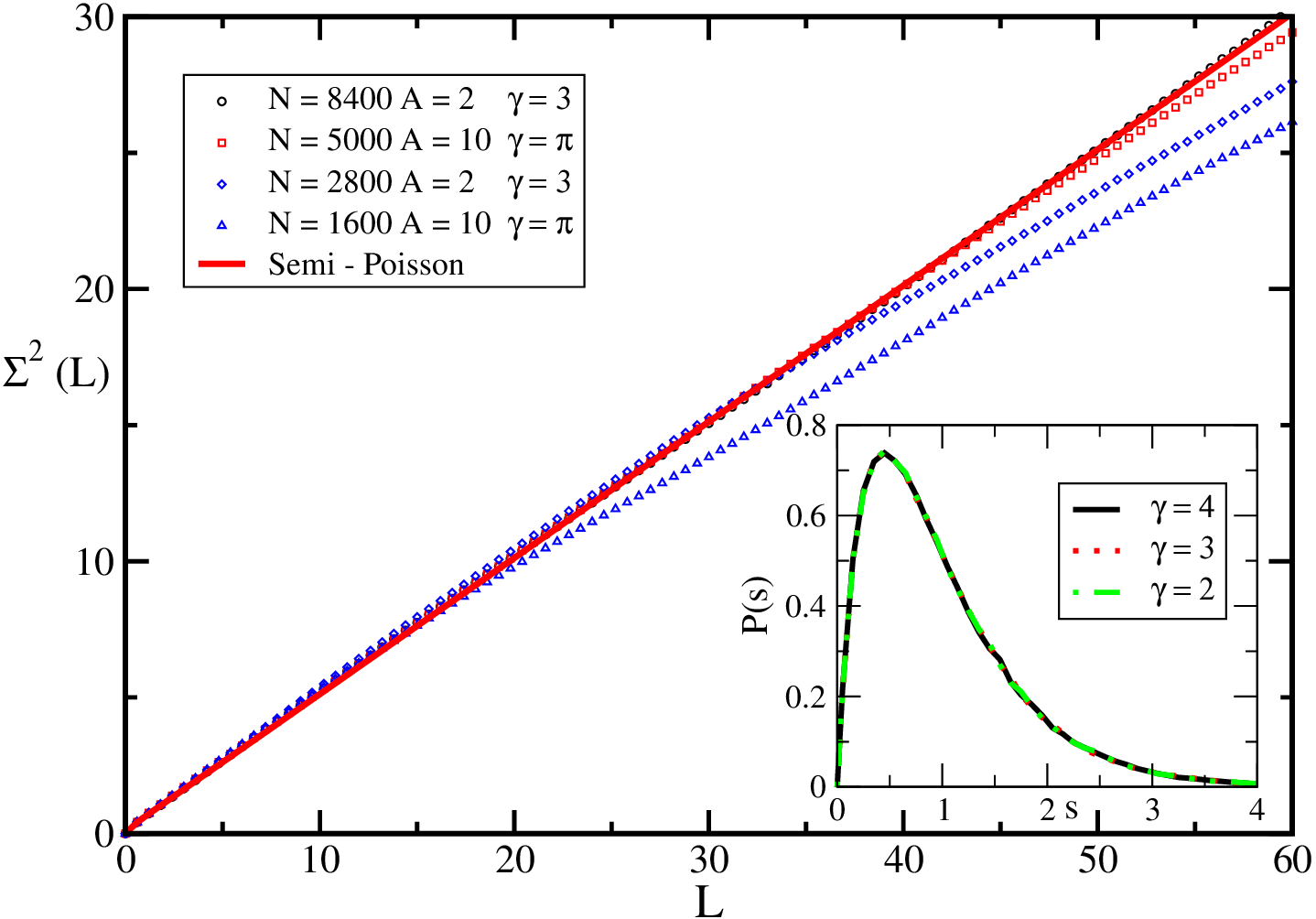}
\vspace{3mm}
\caption{Number variance for different sizes $N$, $A$ and $\gamma$. Provided that 
 $A \gg 1$ the number variance is given by SP statistics for any $N$, $A$ and $\gamma$.
 In the inset $P(s)$ is shown for $A =10$, $N =800$ and different $\gamma$. As shown $P(s)$ is not sensitive to the
 specific value of $\gamma$. }  
\label{fig10} 
\end{figure}

The above analytical arguments have been fully corroborated by numerical calculations.
  By using standard diagonalization techniques we have obtained the eigenvalues of the 
Hamiltonian Eq. (\ref{ourmodel1})
 for different volumes ranging from $N=500$ to $N=8400$. The 
 number of different realizations of disorder is chosen such that
 for each $N$ the total number of eigenvalues be at least $5 \times 10^5$, in all cases $W =1$. 
Eigenvalues close 
to the band edges (around $20 \%$) were discarded from the statistical analysis. 
 The eigenvalues thus obtained 
were unfolded (by using the splines method) with respect to the mean spectral density. 
We first investigate the level statistics in the region $A \gg W$ where, 
according to the analytical findings above, semi-Poisson statistics hold. 
\begin{figure}[hb]
\includegraphics[width=0.7\columnwidth,clip,angle=0]{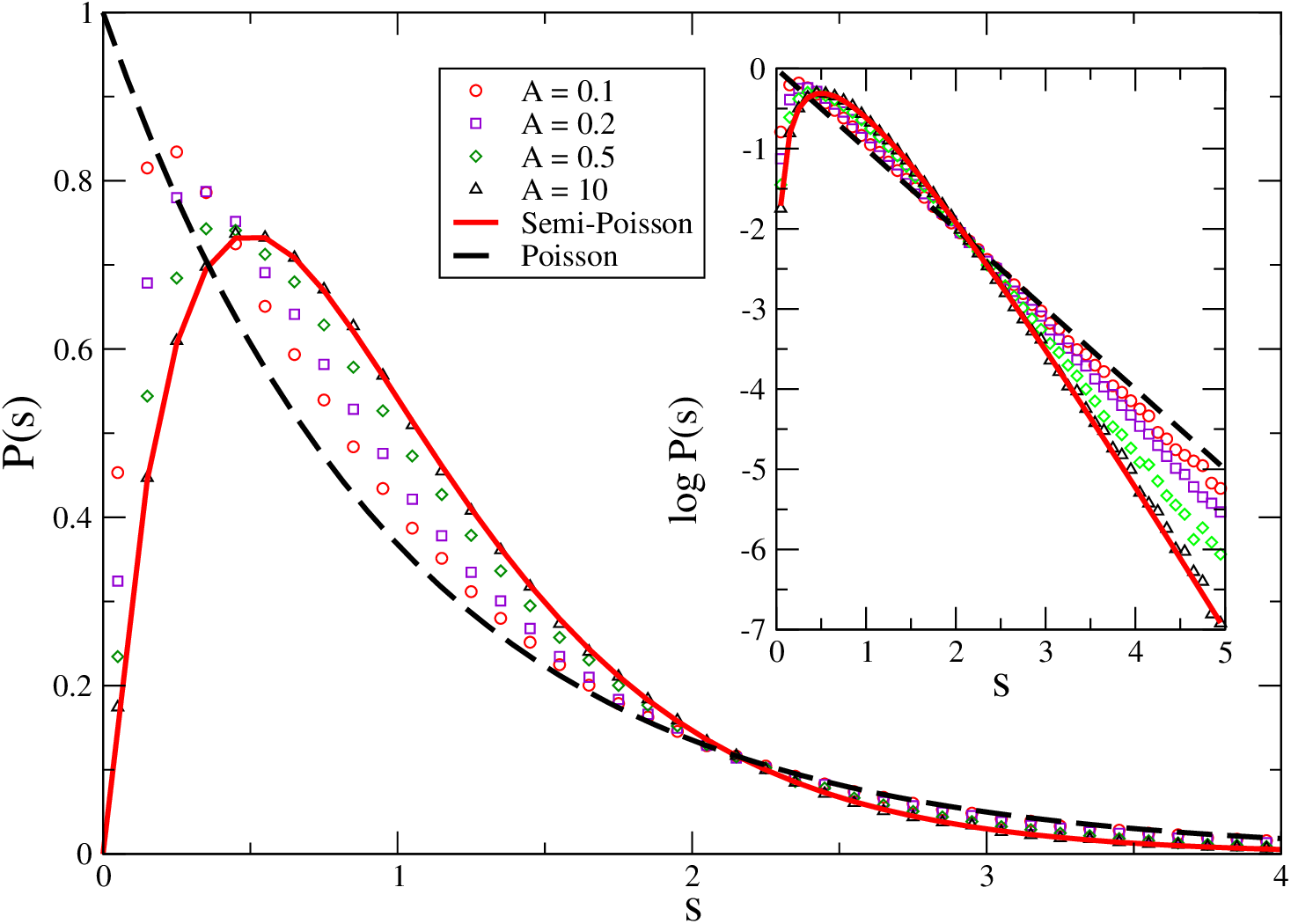}
\vspace{3mm}
\caption{$P(s)$ for $\gamma = \pi/2$, $N =800$ and different $A$. 
A transition to Poisson statistics is observed in the limit 
 $A \ll 1$.}  
\label{fig11} 
\end{figure}  
 As shown in Fig. 1, the 
 level spacing distribution $P(s)$ (including the tail in the inset) 
does not depend on the system size for volumes ranging  
from $N=500$ to $N=8400$. Moreover though level repulsion
$P(s)\propto s~~s\ll 1$ is still present, the asymptotic decay of $P(s)$ (see inset) 
is exponential as for a insulator.
All these features are spectral signatures of an AT.  

 The study of long range correlators as the
 number variance (see Fig. 2), 
$\Sigma^{2}(L)=\langle L \rangle^2 - \langle L^2 \rangle=
\int_{0}^{L}\rho(x)dx -\int_{0}^{L}(L-s)R_{2}(s)ds$, ($\rho(x)$ is the spectral density) 
 further confirms this point. It does not depend on the system size and its asymptotic behavior
 is linear $\Sigma^{2}(L) \sim 0.5 L~~L \gg 1$ as at the AT.

We now compare the level statistics of the Hamiltonian Eq. (\ref{ourmodel1}) with 
 SP.   
 As shown in Fig. 1 and Fig 2 (inset) we could not detect any perceptible deviation from the SP
$P(s)=4se^{-2s}$ prediction for different $\gamma$'s and $A \gg W$. The agreement 
is impressive even for 
 tail of $P(s)$ (inset).  Also long-range correlators as the number variance follow the semi-Poisson prediction $\Sigma_2(L) = L/2 +(1-e^{-4L})/8$ for different parameter values (see Fig.2).
Deviation for small volumes are well known finite size effects. 

A remark is in order, though the above analysis clearly 
show that the level statistics of our model is described by SP
 and share generic features of an 
disordered conductor at the AT there are still important quantitative differences. 
Level statistics at the AT depends on the dimension of the space. For instance in $3D$ ($4D$) the
slope of the number variance  is $0.27$ ($0.41$), by contrast semi-Poisson predicts
$0.5$. Clear differences are also observed in short range correlators as $P(s)$ \cite{varga,ant10} 
where it has been found that, despite their similarities, critical and SP
 have different functional forms. The reason for such discrepancy is as follows:

As mentioned previously, level statistics 
 at the 3D AT (critical statistics) is very accurately described by RMM \cite{Moshe}  whose  joint distribution 
of eigenvalues  
 can be considered as an ensemble of free particles at finite temperature 
with a nontrivial statistical 
 interaction. The statistical interaction resembles
 the Vandermonde determinant, and the effect of finite temperature 
 is to suppress the correlations of distant eigenvalues. 
In SP this suppression is abrupt, in contrast to critical 
statistics, where the effect 
 of the temperature is smooth.
  The reason 
for the differences between SP and critical statistical is thus due to the fact that
 the interaction among eigenvalues is not strictly restricted to nearest neighbors (SP) at the AT.

We now investigate the eigenvector properties of the Hamiltonian Eq. \ref{ourmodel1}. We shall see that, though the 
 they are to some extent multifractals, there exist important difference with respect to those at the AT. 
We have studied the scaling of the eigenfunction moments $P_q$
 with respect to the sample size $L$. For multifractal wavefunctions, 
$\langle P_q \rangle =\int d^dr |\psi({\bf r})|^{2q}\propto L^{-D_q(q-1)}$ where the bracket stands for ensemble average and  $D_q$ is a set of exponents describing the transition.  
Below we show the multifractal dimensions $D_q$ ($\pm 10\%$ ) for $A =10, \gamma = \pi/2$ obtained by numerical fitting of  $\langle \log P_2 \rangle$: $D_{1.5} \sim 0.36, D_2\sim 0.30, D_{2.5} \sim 0.28, D_3 \sim 0.26, D_4 \sim 0.24, D_5 \sim 0.22, D_6 \sim 0.22$.  
For small $q$, $D_q$ depends clearly on $q$ however for larger $q$ the dependence is quite weak suggesting 
that may exist a critical $q_c$ such that for $q > q_c$ 
$D_q \in [0,1]$ is a constant. This situation would correspond with eigenstates which are truly multifractal only up to certain scale. Similar results have been recently reported for certain triangular billiards \cite{bogo04} though in this case it was claimed that $D_q$ is constant for any $q$. We remark that at the 3D AT
 the eigenfunction are truly multifractal and consequently $D_q$ depends explicitly on $q$ for any $q$. 

We now interpret the special properties of eigenfunctions and level statistics in 
the context of the original non-random Hamiltonian consisting of a kick 
rotor with a non-analytical step-like potential. From the above arguments it is 
clear that the Hamiltonian Eq. (\ref{ourmodel}) avoids dynamical localization typical of 
a smooth potential due exclusively 
to the step-like singularity of the potential. 
In a recent paper we found that certain types of classical  singularities
 induce  quantum power-law localization of the corresponding eigenvectors. 
For the case of $\log$ singularities it was explicitly shown that the eigenvector were multifractals and the level statistics
 was given by critical statistics. 
     
The results of this paper shows that SP can be considered as the level statistics associated 
with chaotic systems with classical step-like singularities (in $1+1$ dimensions) or with disordered 
 systems with a step-like dispersion relation. We can thus unify 
different intermediate statistics ('critical statistics' and 'semi-Poisson statistics') in a broader
classification based on the universal relation between classical singularities and level statistics features.

Finally we would like to discuss briefly two different issues. 
 We have observed (see Fig. 3) that as $A$ becomes comparable to $W$ the level statistics shifts slowly toward Poisson. 
The level statistics in this region is still scale invariant and even for $A \ll W$ small deviations from Poisson are not negligible. This finding suggests that the model is indeed critical for all values of $\gamma$ and $A$. Another issue of interest is the robustness of our results under perturbations. 
We have added a flux to the Hamiltonian Eq. (\ref{ourmodel1}) in order to check whether the 
 breaking of time reversal invariance has any impact on the level statistics. 
The results are negatives, we have only observed the effect of the flux in the $s \rightarrow 0$ limit of the level spacing distribution $P(s) \sim s^2$ instead $P(s) \sim s$.  


In conclusion we have introduced a new class of systems with level 
statistics described by semi-Poisson statistics and multifractal wavefunctions.  
The appearance of semi-Poisson 
statistics has been related to a step-like singularities of the classical potential and 
to a singular step-like dispersion relation in a disordered system.  We have discussed similarities
 and differences with critical statistics and claimed that both are part of a larger classification scheme. 
 Finally we have discussed the 
transition to Poisson in our model and the effect of a flux on the level statistics.\\

I thank E. Bogomolny for suggesting me to look at the properties of 
the Hamiltonian Eq. (\ref{ourmodel1}). I acknowledge financial support from the Spanish Ministry 
 of Education and Culture.

\end{document}